\documentclass[11pt,a4paper]{article}
\usepackage{indentfirst}
\usepackage{amsmath,amsthm,amssymb}
\usepackage{graphicx}
\newtheorem{Example}{Example}

\newtheorem{Lemma}{Lemma}

\newcommand{\F}{\mathbb{F}}

\begin{document}

\begin{center} \Large
New Repair strategy of Hadamard Minimum Storage Regenerating Code for Distributed Storage System
\end{center}

\begin{center}
Xiaohu Tang, \emph{Member, IEEE}, Bin Yang, and  Jie Li
\end{center}

\footnote{The authors are with the Information Security and National Computing Grid Laboratory, Southwest Jiaotong University, Chengdu, 610031, China (e-mail: xhutang@swjtu.edu.cn, metroyb@gmail.com, jieli873@gmail.com).}


\textbf{Abstract}--- The newly presented $(k+2,k)$ Hadamard minimum storage regenerating (MSR) code is the first class of high rate storage code with optimal repair property for all single node failures. In this paper, we propose a new simple repair strategy, which can considerably reduces the computation load of the node repair in contrast to the original one.

\normalfont

\section{Introduction}

In distributed storage systems, data is placed on a number of
storage nodes with redundancy. Redundancy is the basis for
distributed storage systems to provide reliable access service.
Normally, there are two mechanisms  of redundancy: replication and
erasure coding. Compared with replication, erasure coding is
becoming more and more attractive because of much better storage
efficiency. Up to now, some famous storage applications, such as
Google Colossus (GFS2) \cite{gfs2}, Microsoft Azure \cite{azure},
HDFS Raid \cite{hdfs-raid}, and OceanStore \cite{oceanstore}, have
adopted erasure coding.

Due to the unreliability of individual storage nodes, node repair will be launched once node failures take place, so as to retain the same redundancy. With data growing much faster than before, node repair becomes a regular maintenance operation now. In general, there are several metrics to evaluate the cost of node repair, such as disk I/O, network bandwidth, number of accessed disks, etc. Among these metrics, the repair bandwidth, defined as the amount of data downloaded to repair a failed node, is the most useful. In \cite{coding}, Dimakis \textit{et al.} established a tradeoff between the storage and repair bandwidth where MBR (minimum bandwidth regenerating) code corresponding to minimum repair bandwidth and MSR (minimum storage regenerating) code corresponding to minimum storage are the most important.

In this study, we focus on  MSR codes with high rate.  So far, several explicit constructions of such MSR codes have been proposed
based on the interference alignment technique \cite{hadamard,zigzag,longMDS}. However, it should be noted that
in all the aforementioned constructions except the one in  \cite{hadamard}, only the systematic nodes possess the optimal repair property.
In \cite{hadamard}, the first $(k,k+2)$ MSR code with optimal repair property for all storage nodes, including both $k$ systematic nodes and $2$ parity nodes, was presented. Actually, the optimal repair property follows from  Hadamard design with the help of lattice representation of the symbol extension technique. Therefore, we call this code Hadamard MSR code throughout this paper.

In this paper, we  fully explore the fundamental properties of Hadamard design. As a result, we present a  generic repair strategy for Hadamard MSR code only based on the elementary mathematics instead of the lattice knowledge. Further, the new generic repair strategy not only includes the original repair strategy
in \cite{hadamard}, but also generates a much more simple but efficient one which can greatly reduce the computation load during the repair of failed nodes.

The remainder of this paper is organized as follows. In Section \ref{section_of_model}, the $(k+2,k)$ Hadamard MSR code is briefly reviewed. In Section \ref{section_of_property}, some fundamental  properties of Hadamard deign are studied to  help the optimal repair. In Section \ref{section_repair_strategy}, the new repair strategy is proposed for systematic nodes, the first parity node and the second parity node respectively. The comparison of computation load between the original strategy in \cite{hadamard} and ours is given in Section \ref{section_of_comparison}. Finally, Section \ref{section_of_conclusion} concludes this paper.

\section{$(k+2,k)$ Hadamard MSR code}\label{section_of_model}

 The $(k+2,k)$ MSR code, consisting of $k$ systematic nodes and $2$ parity nodes, is a typical high rate storage code in distributed storage
 system.  Assume that the original data is of size $M=kN$, it can be equally partitioned into $k$ parts $\textbf{f}=[\textbf{f}_1^T,\textbf{f}_2^T,\cdots,\textbf{f}_k^T]^T$ and placed on $k$ systematic nodes, where $\textbf{f}_i$ is a $N\times 1$ vector.
  In general, 2 parity nodes hold parity data, namely two $N\times 1$ vectors $\textbf{f}_{k+1}$ and $\textbf{f}_{k+2}$,  of all the systematic nodes. Table 1 illustrates the structure of a $(k+2,k)$ MSR code.
\begin{table*}[htbp]\label{Hadamard_Model}
\begin{center}
\caption{Structure of a $(k+2,k)$ MSR code}
\begin{tabular}{|c|c|}
\hline
Systematic node & Systematic data \\
\hline
1 & $\textbf{f}_1$ \\
\hline
\vdots & \vdots \\
\hline
$k$ & $\textbf{f}_k$ \\
\hline
Parity node & Parity data \\
\hline
1 & $\textbf{f}_{k+1}=\textbf{f}_1+ \cdots+ \textbf{f}_k$ \\
\hline
2 & $\textbf{f}_{k+2}=A_1\textbf{f}_1+\cdots + A_k\textbf{f}_k$ \\
\hline
\end{tabular}
\end{center}
\end{table*}

Let $N=2^{k+1}$. The $(k+2,k)$ Hadamard MSR code \cite{hadamard} is characterized by the coding matrices $A_1,\cdots,A_k$ over finite field $\mathbb{F}_q ~(q\ge{2k+3})$ as
\begin{eqnarray}
A_i &=& a_iX_i+b_iX_0+I_N, ~1\le i\le k \nonumber\\
X_j&=&\textrm{diag}(\underbrace{I_{2^j},-I_{2^j},\cdots,I_{2^j},-I_{2^j}}\limits_{2^{k+1-j}}),\label{A_i_definition-2}
~0\le j\le k,
\end{eqnarray}
where $I_m$ is the identity matrix of order $m$, the elements $a_i\ne 0$ and $b_i\ne 0$ over the finite field of odd characteristic and order $q\ge 2k+3$ satisfy
\begin{eqnarray}
a_i^2-b_i^2&=&-1,\label{Eqn_a-req-1}\\
a_i\pm a_j&\ne&  b_i- b_j,\nonumber\\
a_i\pm a_i&\ne&  -(b_i- b_j),\nonumber
\end{eqnarray}
for all $1\le i\ne j\le k$ \cite{hadamard}.
In fact, the  matrices in \eqref{A_i_definition-2} are  built on Hadamard design \cite{DS92}.

As the same as other $(k+2,k)$ MSR codes, this $(k+2,k)$ Hadamard
MSR code can tolerate $2$ arbitrary node failures \cite{hadamard}.
Notably, recall that this Hadamard MSR code has an advantage over
other $(k+2,k)$ MSR codes that both systematic nodes and parity
nodes have optimal repair property. Indeed, to repair a failed node
$1\le i\le k+2$, the optimal repair property requires  downloading
$N/2=2^{k}$ data from each surviving node $1\le l\ne i\le k+2$ by
multiplying its original data $\textbf{f}_l$ with a $N/2\times N$ matrix \cite{hadamard},
which will be discussed in detail in Section \ref{section_repair_strategy}.

\begin{Example}\label{Exm_1}
For $k=2$, the $(4,2)$ Hadamard MSR code has the following coding matrices over $\mathbb{F}_{7}$
\begin{eqnarray*}
A_1 &=& \mathrm{diag}(1,1,-1,-1,1,1,-1,-1)+3\cdot\mathrm{diag}(1,-1,1,-1,1,-1,1,-1)+I_{8}\\
A_2 &=& \mathrm{diag}(1,1,1,1,-1,-1,-1,-1)+4\cdot\mathrm{diag}(1,-1,1,-1,1,-1,1,-1)+I_{8}
\end{eqnarray*}
Its repair matrices will be elaborated in Section \ref{section_repair_strategy}.
\end{Example}

\begin{Example}\label{Exm_2}
For $k=3$, the $(5,3)$ Hadamard MSR code has the following coding
matrices over $\mathbb{F}_{11}$
\begin{eqnarray*}
A_1 &=& 2\cdot\mathrm{diag}(1,1,-1,-1,1,1,-1,-1,1,1,-1,-1,1,1,-1,-1)+\\
&&7\cdot\mathrm{diag}(1,-1,1,-1,1,-1,1,-1,1,-1,1,-1,1,-1,1,-1)+I_{16}\\
A_2 &=& 2\cdot\mathrm{diag}(1,1,1,1,-1,-1,-1,-1,1,1,1,1,-1,-1,-1,-1)+\\
&&4\cdot\mathrm{diag}(1,-1,1,-1,1,-1,1,-1,1,-1,1,-1,1,-1,1,-1)+I_{16}\\
A_3 &=& 6\cdot\mathrm{diag}(1,1,1,1,1,1,1,1,-1,-1,-1,-1,-1,-1,-1,-1)+\\
&&2\cdot\mathrm{diag}(1,-1,1,-1,1,-1,1,-1,1,-1,1,-1,1,-1,1,-1)+I_{16}
\end{eqnarray*}
\end{Example}

\section{Properties about Hadamard design}\label{section_of_property}

For $0\le i\le k$, to characterize the diagonal matrix $X_i$ in \eqref{A_i_definition-2} from Hadamard design,
we define $\textbf{x}_i=(x^i_j)_{j=0}^{N-1}$ to be the row vector of length $N$ formed by
its elements of the main diagonal, i.e.,
\begin{eqnarray*}\label{Eqn_Xi}
\textbf{x}_i=(\underbrace{\textbf{1}_{2^i},-\textbf{1}_{2^i},\cdots,\textbf{1}_{2^i},-\textbf{1}_{2^i}}\limits_{2^{k+1-i}})
\end{eqnarray*}
where $\textbf{1}_{2^i}$ is the all one row vector of length $2^{i}$. For example, when $k=2$,
\begin{eqnarray*}
\textbf{x}_0 &=& (1,-1,1,-1,1,-1,1,-1)\\
\textbf{x}_1 &=& (1,1,-1,-1,1,1,-1,-1)\\
\textbf{x}_2 &=& (1,1,1,1,-1,-1,-1,-1)
\end{eqnarray*}

 The following properties of $\textbf{x}_i$
are obvious:
\begin{itemize}
\item \textbf{Alternative Property}: $x_j^i=-x_{j+2^i}^i$ for $0\le j< N-2^i$;
\item \textbf{Periodic Property}: $x_j^i=x_{j+2^{i+1}}^i$ for $0\le j< N-2^{i+1}$, i.e., $\textrm{x}_i$ has period $2^{i+1}$;
\item \textbf{Run Property}: $x_j^i=(-1)^{\lfloor{j/{2^i}}\rfloor}$ for $0\le j< N$, i.e., $\textrm{x}_i$ has $2^{k+1-i}$ runs of
$1$ or $-1$ of length $2^i$;
\item \textbf{Skew-symmetric Property}: $x^i_j=-x^i_{N-1-j}$ for $0\le j< N$.
\end{itemize}

Based on the above properties, we derive the following useful lemmas, which
are crucial to our  repair strategy.
\begin{Lemma}\label{lem_pro}
For any $0\le i, l\le k$,
$j=\mu 2^{l+1}+\nu$,  $0\le \mu<2^{k-l}$, and $0\le \nu<{2^l}$,
\begin{eqnarray}\label{Eqn_Had}
x_j^i=\left\{\begin{array}{rl}
-x_{j+2^{l}}^i, & i=l\\
x_{j+2^{l}}^i, &\textrm{otherwise}
\end{array}
\right.
\end{eqnarray}
\end{Lemma}
\textit{Proof}: Firstly, when $i=l$, \eqref{Eqn_Had} holds due to the alternative property.
Secondly, when $i<l$, \eqref{Eqn_Had} is true because of  the periodic property.
Thirdly, when $i>l$, write $\mu=\mu_0 2^{i-l-1}+\mu_1$ where $0\le\mu_0<2^{k+1-i}$ and $0\le \mu_1\le 2^{i-l-1}-1$, then
\begin{eqnarray*}
\lfloor {j\over {2^i}}\rfloor=\lfloor { j+2^l\over 2^i} \rfloor=\mu_0
\end{eqnarray*}
since $0\le \mu_1 2^{l+1}+2^l+\nu\le 2^i-2^{l+1}+2^{l}+\nu <2^i$, which results in \eqref{Eqn_Had} by the run property.
\hfill$\Box$

\begin{Lemma}\label{lem_near_skew_symmetry}
For any $0\le i\le k$ and $0\le j<N/2$,
\begin{eqnarray*}
x_{N-1-j-(-1)^j}^i =\left\{
\begin{array}{rl}
x_j^i, & i=0\\
-x_j^i, & 0<i\le k
\end{array}
\right.
\end{eqnarray*}
\end{Lemma}
\textit{Proof}:  When $i=0$, the result directly follows from the periodic property that $\textrm{x}_0$ has period $2$ and
$2|(N-1-2j-(-1)^j)$.

When $0<i\le k$, let $j=\mu 2^i+\nu$ where $0\le\mu<2^{k+1-i}$ and $0\le \nu<2^{i}$. According to the run property,
$x_j^i=(-1)^{\mu}$ and
\begin{eqnarray*}
x_{N-1-j-(-1)^j}^i={(-1)}^{\lfloor {N-1-j-(-1)^j\over 2^i} \rfloor}=(-1)^{2^{k+1-i}-\lceil {1+j+(-1)^j\over 2^i}\rceil}
=(-1)^{\lceil {1+j+(-1)^j\over 2^i}\rceil}
\end{eqnarray*}
If $j$ is even, $1+j+(-1)^j=j+2=\mu 2^i+\nu+2$, which implies $\lceil {1+j+(-1)^j\over 2^i}\rceil=\mu+1$ since $0\le \nu\le 2^i-2$ in this case.
If $j$ is odd, $1+j+(-1)^j=j=\mu 2^i+\nu$, which still gives $\lceil {1+j+(-1)^j\over 2^i}\rceil=\mu+1$ since $1\le \nu\le 2^i-1$. Therefore we always have
\begin{eqnarray*}
x_{N-1-j-(-1)^j}^i={(-1)}^{\mu+1}=-x_j^i
\end{eqnarray*}
\hfill$\Box$

Sylvester Hadamard matrices are one of the earliest infinite family of
Hadamard matrices recursively defined by
\begin{equation*}
H_1= \left(
\begin{array}{cc}
1 & 1\\
1 & -1
\end{array}
\right)
\end{equation*}
and
\begin{equation}\label{Eqn_SyH}
H_k= \left(
\begin{array}{cc}
H_{k-1} & H_{k-1}\\
H_{k-1} & -H_{k-1}
\end{array}
\right), ~k\ge 2.
\end{equation}

Normally,  when a $2^k\times 2^k$ matrix, with each entry being $1$ or $-1$, is multiplied by a column vector of length $2^k$,
we do not need multiplication and what we  need are $2^k(2^k-1)$ additions. But for the Sylvester Hadamard
matrix, we can reduce the number of  additions by means of the recursive property.

\begin{Lemma}\label{lem_H_multiply_f}
Let $H_k$ be the Sylvester Hadamard matrix in \eqref{Eqn_SyH} and $\mathbf{z}$ be an arbitrary column vector  of length $2^k$
where $k$ is a positive integer. Then,
\begin{enumerate}
\item [(1)] To compute $H_k\cdot \mathbf{z}$, $k\cdot 2^k$ additions are needed;
\item [(2)] To compute $(H_{k-1}~H_{k-1})\mathbf{z}$ or $(H_{k-1}~-H_{k-1})\mathbf{z}$, $k 2^k-2^{k-1}$ additions are needed.
\end{enumerate}
\end{Lemma}
\textit{Proof}:    Let $\mathcal{N}_k$ denote the number of
additions of $H_k\cdot \mathbf{z}$.

(1) We prove the first assertion by induction. Obviously, it is true for $k=1$, i.e., $\mathcal{N}_1=2$.

Note that
\begin{eqnarray}\label{Eqn_Hadamard_Comp}
H_k\textbf{z} &=& \left(
\begin{array}{cc}
H_{k-1} & H_{k-1} \\
H_{k-1} & -H_{k-1}
\end{array}
\right)
\left(
\begin{array}{c}
\textbf{z}^1 \\
\textbf{z}^2
\end{array}
\right)\nonumber\\
&=&
\left(
\begin{array}{c}
H_{k-1}\textbf{z}^1+H_{k-1}\textbf{z}^2 \\
H_{k-1}\textbf{z}^1-H_{k-1}\textbf{z}^2
\end{array}
\right)
\end{eqnarray}
where $\textbf{z}^1$ and $\textbf{z}^2$ are two column vectors  of length $2^{k-1}$.
Then, we have
\begin{eqnarray*}
\mathcal{N}_k=2 \mathcal{N}_{k-1}+2^k=2^{k-1} \mathcal{N}_1+(k-1)2^k=k\cdot 2^k.
\end{eqnarray*}

(2) The second assertion follows directly from
\eqref{Eqn_Hadamard_Comp}.

\hfill$\Box$

\section{Optimal repair strategy}\label{section_repair_strategy}

Let $\{\textbf{e}_0,\cdots,\textbf{e}_{2^k-1}\}$ be the basis of
$\mathbb{F}_q^{2^k}$. For example, it can be simply chosen as the standard basis
\begin{equation}\label{Eqn_Standard_Basis}
    \textbf{e}_i=(\underbrace{0,\cdots,0,1,0,\cdots,0}\limits_{2^k})^T
\end{equation}
with only the $i$th entry being nonzero.

In this section, we  present our repair strategy respectively for
the systematic nodes, the first parity node, and the second parity
node by giving the corresponding repair matrices, and then check the
optimality.

\subsection{Optimal repair of systematic nodes}\label{subsection_repair_systematic_node}

In order to repair the $i$th systematic node, $1\le i\le k$,  one
downloads data $S_i \textbf{f}_l$, $1\le l\ne i\le k+2$, where the
$N/2\times N$ repair matrix $S_i$ is
\begin{eqnarray*}\label{systematic_node_repair_matrix_element}
S_i=
(\underbrace{\textbf{e}_0,\cdots,\textbf{e}_{2^i-1}}\limits_{2^i},\underbrace{\textbf{e}_0,\cdots,\textbf{e}_{2^i-1}}\limits_{2^i},\cdots,
\underbrace{\textbf{e}_{2^k-2^i},\cdots,\textbf{e}_{2^k-1}}\limits_{2^i},\underbrace{\textbf{e}_{2^k-2^i},\cdots,\textbf{e}_{2^k-1}}\limits_{2^i})
\end{eqnarray*}
Let $s^i_j$ be the $j$th column vector of $S_i$. Obviously,
$\textbf{s}_j^i=\textbf{e}_{\mu 2^i+\nu}$ and
\begin{eqnarray}\label{systematic_node_repair_matrix_element-1}
\textbf{s}_{j+2^i}^i=\textbf{s}_j^i
\end{eqnarray}
where $j=\mu 2^{i+1}+\nu$,  $0\le\mu<2^{k-i}$ and $0\le \nu<2^{i}$.

Then, the data from two parity nodes are
\begin{eqnarray}\label{Eqn_Sys_IF}
\left(
\begin{array}{c}
S_i \\
S_iA_i
\end{array}
\right)\textbf{f}_i + \sum_{l=1,l\ne i}^k \left( \begin{array}{c}
S_i \\
S_iA_l
\end{array}
\right)\textbf{f}_l
\end{eqnarray}
where the second term is the interference resulted from systematic
nodes except the failed one. To cancel the interference and recover
the data $\textbf{f}_i$, the optimal repair strategy requires
\cite{hadamard}
\begin{eqnarray}\label{repair_systematic_node_requirement1}
\textrm{rank} \left(
\begin{array}{c}
S_i \\
S_iA_i
\end{array}
\right) = N
\end{eqnarray}
and
\begin{eqnarray}\label{repair_systematic_node_requirement2}
\textrm{rank} \left(
\begin{array}{c}
S_i \\
S_iA_l
\end{array}
\right) = {N\over 2}
\end{eqnarray}
for $1\le i\ne l\le k$.

Multiplying $A_l$ by $S_i$, $1\le l\le k$, we get
\begin{eqnarray}\label{Eqn_Sysm-1}
S_i A_l=((a_lx_0^l+b_l x_0^0+1)\textbf{s}_0^i~\cdots~(a_lx_j^l+b_l x_j^0+1)\textbf{s}_j^i~\cdots~(a_lx_{N-1}^l+b_{l} x_{N-1}^0+1)\textbf{s}_{N-1}^i)
\end{eqnarray}
Consider the  submatrix of $\left(
\begin{array}{c}
S_i \\
S_iA_l
\end{array}
\right)$ formed by
columns $j$ and $j+2^i$ where $j=\mu 2^{i+1}+\nu$, $0\le \mu< 2^{k-i}$ and $0\le \nu<2^i$, i.e.,
\begin{eqnarray*}\label{Eqn_Sysm}
\Delta_j=\left(
\begin{array}{cc}
\textbf{s}^i_j & \textbf{s}^i_{j+2^i}\\
(a_lx_j^l+b_l x_j^0+1)\textbf{s}^i_j & (a_lx_{j+2^i}^l+b_{l} x_{j+2^i}^0+1)\textbf{s}^i_{j+2^i}
\end{array}
\right)
\end{eqnarray*}
By Lemma \ref{lem_pro}, \eqref{Eqn_a-req-1} and \eqref{systematic_node_repair_matrix_element-1}, we then have
\begin{eqnarray*}
\textrm{rank}
(\Delta_j)
=
\left\{
\begin{array}{ll}
2, & \textrm{if}~i=l\\
1, & \textrm{otherwise}
\end{array}
\right.
\end{eqnarray*}
which results in \eqref{repair_systematic_node_requirement1} and
\eqref{repair_systematic_node_requirement2}.

\begin{Example} When $k=2$, for the $(4,2)$ Hadamard MSR code determined by the coding matrices given in
Example \ref{Exm_1}, the repair matrices of systematic nodes 1 and 2
are respectively
\begin{eqnarray*}
S_1=
\left(
\begin{array}{llllllll}
1 & 0 & 1 & 0 & 0 & 0 & 0 & 0\\
0 & 1 & 0 & 1 & 0 & 0 & 0 & 0\\
0 & 0 & 0 & 0 & 1 & 0 & 1 & 0\\
0 & 0 & 0 & 0 & 0 & 1 & 0 & 1
\end{array}
\right),
S_2=
\left(
\begin{array}{cccccccc}
1 & 0 & 0 & 0 & 1 & 0 & 0 & 0\\
0 & 1 & 0 & 0 & 0 & 1 & 0 & 0\\
0 & 0 & 1 & 0 & 0 & 0 & 1 & 0\\
0 & 0 & 0 & 1 & 0 & 0 & 0 & 1
\end{array}
\right)
\end{eqnarray*}
\end{Example}

\subsection{Optimal repair of the first parity node}\label{first_parity_node_repair_subsection}

In order to repair the first parity node, we need the following transformation
\begin{eqnarray*}
\textbf{y}_1&=&\textbf{f}_1+\cdots+\textbf{f}_k\\
\textbf{y}_i&=&-\textbf{f}_i,~~2\le i\le k
\end{eqnarray*}
Let $\textbf{y}=[\textbf{y}_1^T,\cdots,\textbf{y}_k^T]^T$. The storage code can then be described as
\begin{eqnarray*}
\left(
\begin{array}{c}
\textbf{f}_{k+1}\\
-\textbf{f}_2\\
\vdots\\
-\textbf{f}_{k}\\
\textbf{f}_1\\
\textbf{f}_{k+2}
\end{array}
\right)
=\left(
\begin{array}{cccc}
I_N & 0_N & \cdots & 0_N \\
0_N & I_N & \cdots & 0_N \\
\vdots & \vdots & \ddots & \vdots \\
0_N & 0_N & \cdots & I_N \\
I_N & I_N & \cdots & I_N \\
A_1 & A_1-A_2 & \cdots & A_1-A_k
\end{array}
\right)
\cdot
\textbf{y}
\end{eqnarray*}
where  the first systematic node and the first
parity node are exchanged.

Thus, it suffices to repair the new first systematic node by
respectively downloading data $S \textbf{f}_i$, $1\le i\le k$, and
$\tilde{S}\textbf{f}_{k+2}$, where the repair matrices $S$ and $\tilde{S}$
are
\begin{eqnarray*}
S&=&
(\underbrace{\textbf{e}_0,\textbf{e}_1,\cdots,\textbf{e}_{2^k-2},\textbf{e}_{2^k-1}}\limits_{2^k},\underbrace{\textbf{e}_{2^k-1},\textbf{e}_{2^k-2},\cdots,\textbf{e}_1,\textbf{e}_0}\limits_{2^k})\\
\tilde{S}&=&
(\underbrace{\textbf{e}_0,\textbf{e}_1,\cdots,\textbf{e}_{2^k-2},\textbf{e}_{2^k-1}}\limits_{2^k},\underbrace{-\textbf{e}_{2^k-1},-\textbf{e}_{2^k-2},\cdots,-\textbf{e}_1,-\textbf{e}_0}\limits_{2^k})
\end{eqnarray*}
with the $j$th columns $\textbf{s}_j$ and $\tilde{\textbf{s}}_j$ satisfying
\begin{eqnarray}\label{Eqn_1st-Pnode}
\textbf{s}_j&=&\textbf{s}_{N-1-j}\nonumber\\
\tilde{\textbf{s}}_j&=&-\tilde{\textbf{s}}_{N-1-j}
\end{eqnarray}
for $0\le j< N$.

Then, the data from the new first parity node and the second parity
node can be expressed as
\begin{eqnarray*}\label{Eqn_download-1}
\left(
\begin{array}{c}
S \\
\tilde{S}A_1
\end{array}
\right)\textbf{f}_{k+1} - \sum_{l=2}^k \left( \begin{array}{c}
S \\
\tilde{S} (A_1-A_l)
\end{array}
\right)\textbf{f}_l.
\end{eqnarray*}
The optimal repair strategy requires \cite{hadamard}
\begin{eqnarray}\label{repair_1parity_node_requirement1}
\textrm{rank} \left(
\begin{array}{c}
S \\
\tilde{S}A_1
\end{array}
\right) = N
\end{eqnarray}
and
\begin{eqnarray}\label{repair_1parity_node_requirement2}
\textrm{rank} \left(
\begin{array}{c}
S \\
\tilde{S} (A_1-A_l)
\end{array}
\right) = {N\over 2}
\end{eqnarray}
for $2\le l\le k$.

According to \eqref{repair_1parity_node_requirement1} and
\eqref{repair_1parity_node_requirement2}, we investigate
\begin{eqnarray*}
\left(\begin{array}{c}
S\\
\tilde{S} A_1
\end{array}\right)=\left(\begin{array}{ccccc}
\textbf{s}_{0}& \cdots & \textbf{s}_{j} & \cdots &
\textbf{s}_{N-1}\\
(a_1 x_{0}^1+b_{1} x_{0}^0+1)\tilde{\textbf{s}}_{0} & \cdots & (a_1 x_{j}^1+b_{1} x_{j}^0+1)\tilde{\textbf{s}}_{j}&\cdots & (a_1 x_{N-1}^1+b_{1} x_{N-1}^0+1)\tilde{\textbf{s}}_{N-1}\\
\end{array}\right)
\end{eqnarray*}
and
\begin{eqnarray*}
&&\left(\begin{array}{c}
S\\
\tilde{S} (A_1-A_l)
\end{array}\right)\\
&=&\left(\begin{array}{ccc} \textbf{s}_{0}& \cdots & \textbf{s}_{j}\\
(a_1x_{0}^1+(b_{1}-b_{l}) x_{0}^0-a_lx_{0}^l)\textbf{s}_{0} & \cdots & (a_1x_j^1+(b_{1}-b_{l})
x_{j}^0-a_l
x_{j}^l)\tilde{\textbf{s}}_{j}\end{array}\right.\\
&&\left.\begin{array}{cccc}
&&\cdots & \textbf{s}_{N-1}\\
&&\cdots & (a_1
x^1_{N-1}+(b_{1}-b_{l}) x_{N-1}^0-a_l x_{N-1}^l)\tilde{\textbf{s}}_{N-1}
\end{array}\right)
\end{eqnarray*}
The submatrices formed by columns $j$ and $N-1-j$, $0\le j<N/2$,
are respectively
\begin{eqnarray*}
\Delta_j&=&\left(\begin{array}{cc} \textbf{s}_{j} & \textbf{s}_{N-1-j}\\
(a_1 x_{j}^1+b_{1} x_{j}^0+1)\tilde{\textbf{s}}_{j} & (a_1
x_{N-1-j}^1+b_{1} x_{N-1-j}^0+1)\tilde{\textbf{s}}_{N-1-j}
\end{array}\right)\\
&=&\left(\begin{array}{cc} \textbf{s}_{j} & \textbf{s}_{j}\\
(a_1 x_{j}^1+b_{1} x_{j}^0+1)\tilde{\textbf{s}}_{j} & (a_1 x_{j}^1+b_{1}
x_{j}^0-1)\tilde{\textbf{s}}_{j}
\end{array}\right)
\end{eqnarray*}
and
\begin{eqnarray}\label{Eqn_Parity-1}
\Gamma_j&=&\left(\begin{array}{cc} \textbf{s}_{j} & \textbf{s}_{N-1-j}\\
(a_1x_j^1+(b_{1}-b_{l}) x_{j}^0-a_l x_{j}^l)\tilde{\textbf{s}}_{j} & (a_1x_{N-1-j}^1+(b_{1}-b_{l})
x_{N-1-j}^0-a_l
x_{N-1-j}^l)\tilde{\textbf{s}}_{N-1-j}
\end{array}\right)\nonumber\\
&=&\left(\begin{array}{cc} \textbf{s}_{j} & \textbf{s}_{j}\\
(a_1x_j^1+(b_{1}-b_{l}) x_{j}^0-a_l x_{j}^l)\tilde{\textbf{s}}_{j}& (a_1x_j^1+(b_{1}-b_{l}) x_{j}^0-a_l
x_{j}^l)\tilde{\textbf{s}}_{j}
\end{array}\right)
\end{eqnarray}
by the skew-symmetric property and \eqref{Eqn_1st-Pnode}. In other
words,
\begin{eqnarray*}
\textrm{rank}(\Delta_j)=2,~~~\textrm{rank}(\Gamma_j)=1
\end{eqnarray*}
which leads to \eqref{repair_1parity_node_requirement1}
and \eqref{repair_1parity_node_requirement2}.

\begin{Example} When $k=2$, for the $(4,2)$ Hadamard MSR code determined by the coding matrices given in
Example \ref{Exm_1}, the repair matrices of the first parity node
are
\begin{eqnarray*}
S=
\left(
\begin{array}{cccccccc}
1 & 0 & 0 & 0 & 0 & 0 & 0 & 1\\
0 & 1 & 0 & 0 & 0 & 0 & 1 & 0\\
0 & 0 & 1 & 0 & 0 & 1 & 0 & 0\\
0 & 0 & 0 & 1 & 1 & 0 & 0 & 0\\
\end{array}
\right),
\tilde{S}=
\left(
\begin{array}{cccccccc}
1 & 0 & 0 & 0 & 0 & 0 & 0 & -1\\
0 & 1 & 0 & 0 & 0 & 0 & -1 & 0\\
0 & 0 & 1 & 0 & 0 & -1 & 0 & 0\\
0 & 0 & 0 & 1 & -1 & 0 & 0 & 0\\
\end{array}
\right)
\end{eqnarray*}
\end{Example}

\subsection{Optimal repair of the second parity node}

Similar to the repair of the first parity node, the second parity node can be regarded as  the first systematic node by the following transformation
\begin{eqnarray*}
\textbf{y}_1&=&A_1\textbf{f}_1+\cdots+A_k\textbf{f}_k \\
\textbf{y}_i&=&-A_i \textbf{f}_i
,~2\le i\le k
\end{eqnarray*}
Let $\textbf{y}=[\textbf{y}_1^T,\cdots,\textbf{y}_k^T]^T$. With this transformation, the storage code can be described as
\begin{eqnarray*}
\left(
\begin{array}{c}
\textbf{f}_{k+2}\\
-A_2\textbf{f}_2\\
\vdots\\
-A_kf_{k}\\
A_1\textbf{f}_{1}\\
\textbf{f}_{k+1}
\end{array}
\right)=\left(
\begin{array}{cccc}
I_N & 0_N & \cdots & 0_N \\
0_N & I_N & \cdots & 0_N \\
\vdots & \vdots & \ddots & \vdots \\
0_N & 0_N & \cdots & I_N \\
I_N & I_N & \cdots & I_N \\
A_1^{-1} & A_1^{-1}-A_2^{-1} & \cdots & A_1^{-1}-A_k^{-1}
\end{array}
\right)
\cdot
\textbf{y}
\end{eqnarray*}
where the three nodes, i.e., the first systematic node, the first  parity node and the second parity node, are cyclically shifted.

Hence, it is sufficient to repair the new first systematic node by
downloading data $S A_i\textbf{f}_i$, $1\le i\le k$, and
$\tilde{S}\textbf{f}_{k+1}$, where the two repair matrices $S$ and
$\tilde{S}$ are
\begin{eqnarray*}
S&=&
(\underbrace{\textbf{e}_0,\textbf{e}_1,\cdots,\textbf{e}_{2^k-2},\textbf{e}_{2^k-1}}\limits_{2^k},\underbrace{\textbf{e}_{2^k-2},\textbf{e}_{2^k-1},\cdots,\textbf{e}_0,\textbf{e}_1}\limits_{2^k})\\
\tilde{S}&=&
(\underbrace{\textbf{e}_0,\textbf{e}_1,\cdots,\textbf{e}_{2^k-2},\textbf{e}_{2^k-1}}\limits_{2^k},\underbrace{-\textbf{e}_{2^k-2},-\textbf{e}_{2^k-1},\cdots,-\textbf{e}_0,-\textbf{e}_1}\limits_{2^k})
\end{eqnarray*}
with the $j$th columns  $\textbf{s}_j$ and $\tilde{\textbf{s}}_j$ being
\begin{eqnarray*}
\textbf{s}_j&=&
\left\{
\begin{array}{ll}
\textbf{e}_j,&0\le j<N/2\\
\textbf{e}_{N-1-j-(-1)^j}, &N/2 \le j< N
\end{array}
\right.
\\
\tilde{\textbf{s}}_j&=&
\left\{
\begin{array}{ll}
\textbf{e}_j,&0\le j<N/2\\
-\textbf{e}_{N-1-j-(-1)^j}, &N/2 \le j< N
\end{array}
\right.
\end{eqnarray*}
satisfying
\begin{eqnarray}\label{first_parity_node_repair_matrix_property}
\begin{array}{ccc}
\textbf{s}_j & = & \textbf{s}_{N-1-j-(-1)^j}\\
\tilde{\textbf{s}}_j & = & -\tilde{\textbf{s}}_{N-1-j-(-1)^j}
\end{array}
\end{eqnarray}
for $0\le j<N/2$.

Then, the data from the  new first parity node and the new second parity
node can be expressed as
\begin{eqnarray*}\label{Eqn_download-1}
\left(
\begin{array}{c}
S \\
\tilde{S}A_1^{-1}
\end{array}
\right)\textbf{f}_{k+2} - \sum_{l=2}^k \left( \begin{array}{c}
S \\
\tilde{S} (A_1^{-1}-A_l^{-1})
\end{array}
\right)A_l\textbf{f}_l
\end{eqnarray*}
The optimal repair strategy requires \cite{hadamard}
\begin{eqnarray}\label{repair_2parity_node_requirement1}
\textrm{rank} \left(
\begin{array}{c}
S \\
\tilde{S}A_1^{-1}
\end{array}
\right) = N
\end{eqnarray}
and
\begin{eqnarray}\label{repair_2parity_node_requirement2}
\textrm{rank} \left(
\begin{array}{c}
S \\
\tilde{S} (A_1^{-1}-A_l^{-1})
\end{array}
\right) = {N\over 2}
\end{eqnarray}
for $2\le l\le k$.

By \eqref{repair_2parity_node_requirement1} and
\eqref{repair_2parity_node_requirement2}, we need to discuss $\left(
\begin{array}{c}
S \\
\tilde{S}A_1^{-1}
\end{array}
\right)$
and
$\left(
\begin{array}{c}
S \\
\tilde{S}(A_1^{-1}-A_l^{-1})
\end{array}
\right)
$
where
\begin{eqnarray*}\label{inverse A_i}
A_i^{-1} &=& 2^{-1}(I_N-a_i^{-1}b_iX_0X_i+a_i^{-1}X_i)\\
&=& 2^{-1}I_N+2^{-1}a_i^{-1}X_i(I_N-b_iX_0)
,~1\le i\le k \nonumber
\end{eqnarray*}
and
\begin{eqnarray*}\label{inverse A_i_subtract_inverse_A_1}
A_1^{-1}-A_l^{-1} &=& 2^{-1}(a_l^{-1}b_lX_0X_l-a_1^{-1}b_1X_0X_1+a_1^{-1}X_1-a_l^{-1}X_l)\\
&=& 2^{-1}a_1^{-1}X_1(I_N-b_1X_0)-2^{-1}a_l^{-1}X_l(I_N-b_lX_0),~2\le l\le k\nonumber
\end{eqnarray*}
according to \cite{hadamard}.
For simplicity of the characterization of the matrices $A_1^{-1}$ and $A_1^{-1}-A_l^{-1}$, we define
\begin{eqnarray*}
p^1_j&=&2^{-1}+2^{-1}a_1^{-1}x^1_j(1-b_1 x^0_j)\\
q^l_j&=&2^{-1}a_1^{-1}x^1_j(1-b_1x^0_j)-2^{-1}a_l^{-1}x^l_j(1-b_lx^0_j)
\end{eqnarray*}
where $1< l\le k$ and $0\le j<{N}$.
By  Lemma \ref{lem_near_skew_symmetry},  we have
\begin{eqnarray*}
p^1_{N-1-j-(-1)^j}
&=&2^{-1}-2^{-1}a_1^{-1}x^1_j(1-b_1x^0_j)\\
&=&-p^1_j+1
\end{eqnarray*}
and
\begin{eqnarray*}
q^l_{N-1-j-(-1)^j}
&=&-q^l_j
\end{eqnarray*}
for $0\le j<N/2$.

For $0\le j<N/2$, consider the submatrices formed by columns $j$ and $N-1-j-(-1)^j$ in matrices $\left(
\begin{array}{c}
S \\
\tilde{S}A_1^{-1}
\end{array}
\right)$
and
$\left(
\begin{array}{c}
S \\
\tilde{S}(A_1^{-1}-A_l^{-1})
\end{array}
\right)
$, i.e.,
\begin{eqnarray*}
\Delta_j
&=&
\left(
\begin{array}{cc}
\textbf{s}_j & \textbf{s}_{N-1-j-(-1)^j}\\
p^1_j\tilde{\textbf{s}}_j & p^1_{N-1-j-(-1)^j}\tilde{\textbf{s}}_{N-1-j-(-1)^j}
\end{array}
\right)\\
&=&
\left(
\begin{array}{cc}
\textbf{s}_j & \textbf{s}_j\\
p^1_j\tilde{\textbf{s}}_j & p^1_j\tilde{\textbf{s}}_j-\tilde{\textbf{s}}_j
\end{array}
\right)
\end{eqnarray*}
and
\begin{eqnarray*}
\Gamma_j
&=&
\left(
\begin{array}{cc}
\textbf{s}_j & \textbf{s}_{N-1-j-(-1)^j}\\
q^l_j\tilde{\textbf{s}}_j & q^l_{N-1-j-(-1)^j}\tilde{\textbf{s}}_{N-1-j-(-1)^j}
\end{array}
\right)\\
&=&\left(
\begin{array}{cc}
\textbf{s}_j & \textbf{s}_j\\
q^l_j\tilde{\textbf{s}}_j & q^l_j\tilde{\textbf{s}}_j
\end{array}\right)
\end{eqnarray*}
That is, $\textrm{rank}(\Delta_j)=2$ and $\textrm{rank}(\Gamma_j)=1$, which gives
\eqref{repair_2parity_node_requirement1} and \eqref{repair_2parity_node_requirement2}.

\begin{Example}When $k=2$, for the $(4,2)$ Hadamard MSR code determined by the coding matrices given in
Example \ref{Exm_1}, the repair matrices of the second parity node
are
\begin{eqnarray*}
S=
\left(
\begin{array}{cccccccc}
1 & 0 & 0 & 0 & 0 & 0 & 1 & 0 \\
0 & 1 & 0 & 0 & 0 & 0 & 0 & 1 \\
0 & 0 & 1 & 0 & 1 & 0 & 0 & 0 \\
0 & 0 & 0 & 1 & 0 & 1 & 0 & 0
\end{array}
\right),
\tilde{S}=
\left(
\begin{array}{cccccccc}
1 & 0 & 0 & 0 & 0 & 0 & -1 & 0 \\
0 & 1 & 0 & 0 & 0 & 0 & 0 & -1 \\
0 & 0 & 1 & 0 & -1 & 0 & 0 & 0 \\
0 & 0 & 0 & 1 & 0 & -1 & 0 & 0
\end{array}
\right)
\end{eqnarray*}
\end{Example}

\section{Comparison}\label{section_of_comparison}

In fact, in the original repair  strategy \cite{hadamard} the basis
$\{\textbf{e}_0,\cdots,\textbf{e}_{2^k-1}\}$ is chosen as the
column vectors of the Sylvester Hadamard matrix in \eqref{Eqn_SyH}.
Whereas, for our strategy,
$\{\textbf{e}_0,\cdots,\textbf{e}_{2^k-1}\}$ can be any basis of
$\mathbb{F}_q^{2^k}$. In this sense, our new repair strategy
generalizes the previous one in \cite{hadamard}.

Most importantly,  by choosing the standard basis in
\eqref{Eqn_Standard_Basis}, our new repair strategy can considerably
reduce the computation, including both addition and multiplication,
in contrast to  the original repair strategy in \cite{hadamard}. Indeed, the
decrease comes  from the fact that in each row, our new  repair
matrices  have  $2$ nonzero elements of $1$ or $-1$ whereas the
original matrices have $N$ nonzero elements of $1$ or $-1$.

The computation of node repair lies in 3 phases: download, interference cancellation and recover. In what follows,
we investigate it case by case.

\vspace{3mm}\textbf{Case 1. Computation load of the repair of systematic nodes}

Since each $S_i \cdot\textbf{f}_l$ needs $N/2$ additions, the new
strategy  needs  $(k+1)N/2$ additions in the download phase. When $i\ne l$, note
that in \eqref{Eqn_Sysm-1} $S_iA_l$ has only two nonzero elements in
each row, which indicates that there exists an $N/2\times N/2$ matrix
\begin{eqnarray}\label{Eqn_simp_comp}
B_l=\textrm{diag}(p^l_0,\cdots,p^l_{N/2-1})
\end{eqnarray}
where $p^{l}_{\mu 2^{i}+\nu}=a_{l}x^l_{\mu 2^{i+1}+\nu}+b_l x^0_{\mu 2^{i+1}+\nu}+1$, $0\le\mu<2^{k-i}$ and
$0\le\nu<2^i$ such that $S_iA_l=B_l S_i$. Hence, the new strategy  needs $(k-1)N$
additions  and at most $(k-1)N/2$ multiplications  to  cancel the
interference term in \eqref{Eqn_Sys_IF}.  In the recover phase,  $N$
additions and at most $2N$ multiplications are needed for the new
strategy since the matrix $\left( \begin{array}{c}
S_i \\
S_iA_i
\end{array}
\right)^{-1}$ still has only two nonzero elements in the each row. Therefore,
totally $(3k+1)N/2$ additions and at most $(k+3)N/2$ multiplications are needed
for the new strategy.

For the original strategy, the download phase requires $(k+1)(2k+1)N/2$
additions by Lemma \ref{lem_H_multiply_f} since $S_i$ is equivalent to
$(H_{k}~H_{k})$ with respect to columns permutation; The interference
cancellation phase at most requires $(k-1)(N/2+1)N/2$ additions and $(k-1)N^2/4$
multiplications; The recover phase requires  $N(N-1)$ additions and at most
 $N^2$ multiplications. Thus,
totally $(k+3)N^2/4+(k^2+2k-1)N$ additions and $(k+3)N^2/4$ multiplications are needed
at most.

\vspace{3mm}\textbf{Case 2. Computation load of the repair of the first parity node}

Similarly to case 1, the new strategy  needs $(3k+1)N/2$ additions and at most $(k+3)N/2$ multiplications  because
(1) $\tilde{S} \cdot\textbf{f}_{k+2}$ needs $N/2$ additions, as the same as $ S\cdot\textbf{f}_i$, $1\le i\le k$;
(2) For $2\le l\le k$ there exists an $N/2\times N/2$ matrix
\begin{eqnarray}\label{Eqn_simp_comp-2}
B_l=\textrm{diag}(a_1x_0^1+(b_{1}-b_{l}) x_{0}^0-a_l x_{0}^l, \cdots,
a_1x_{N/2-1}^1+(b_{1}-b_{l}) x_{N/2-1}^0-a_l x_{N/2-1}^l)
\end{eqnarray}
such that $\tilde{S} (A_1-A_l)=B_l S$ by \eqref{Eqn_Parity-1};
(3) The matrix $\left( \begin{array}{c}
S \\
\tilde{S}A_1
\end{array}
\right)^{-1}$ has only two nonzero elements in the each row.

For the original strategy, $(k+3)N^2/4+(k^2+2k-1)N$ additions and $(k+3)N^2/4$ multiplications are required
at most.

\vspace{3mm}\textbf{Case 3. Computation load of the repair of the second parity node}

During the download phase,
the new strategy needs  $(k+1)N/2$ additions and at most $kN$  multiplications
since (1) $S A_i\cdot\textbf{f}_i$, $1\le i\le k$, needs  $N/2$ additions and at most $N$ multiplications;
(2) $\tilde{S}\cdot \textbf{f}_{k+1}$  needs $N/2$ additions.
In the  interference cancellation phase and recover phase, the computation can be analyzed in
the same fashion as that of Case 1. Hence, totally $(3k+1)N/2$ additions and at most $(3k+3)N/2$ multiplications are needed
for the new strategy.

For the original strategy, $(3k+3)N^2/4+(2k-2)N/2$ additions and $(3k+3)N^2/4$ multiplications are needed
at most.

The above comparison is summarized in Table 2, where ADD and MUL respectively denote
the numbers of addition and multiplication. The exact number of additions and multiplications depends on the concrete values of
$a_l, b_l$, $1\le l\le k$,  and the finite field $\F_q$. For the new strategy, the number of multiplications can be further reduced if
set $a_l\pm b_l=\pm 2$ or $a_1 \pm (b_{1}-b_{l})\pm a_l =\pm 1$ such that there are some $1$ or $-1$ in the diagonal
matrix $B_l$ given by \eqref{Eqn_simp_comp} or \eqref{Eqn_simp_comp-2},  which is feasible by the equations (81) and (82)
in \cite{hadamard}. As for the old strategy, it seems hard to be analyzed because
there are too many nonzeros in the Sylvester Hadamard matrix.

\newcommand{\rb}[1]{\raisebox{1.5ex}[0pt]{#1}}

\begin{table*}[htbp]\label{Table_Comparison}
\begin{center}
\caption{Comparison between the original and new strategies for $(k,k+2)$ Hadamard MSR code}
\begin{tabular}{|c|c|c|c|}
\hline
Node & Repair & & \\
to repair & strategy & \rb{ADD} & \rb{MUL}\\
\hline
Systematic & New & $(3k+1)N/2$ & $\le (k+3)N/2$\\
\cline{2-4}
node & Original & $\le (k+3)N^2/4+(k^2+2k-1)N$ & $\le (k+3)N^2/4$ \\
\hline
Parity & New & $(3k+1)N/2$ & $\le (k+3)N/2$ \\
\cline{2-4}
node 1 & Original & $\le (k+3)N^2/4+(k^2+2k-1)N$ & $\le (k+3)N^2/4$\\
\hline
Parity & New & $(3k+1)N/2$ & $\le (3k+3)N/2$ \\
\cline{2-4}
node 2 & Original & $\le (3k+3)N^2/4+(2k-2)N/2$ & $\le (3k+3)N^2/4$\\
\hline
\end{tabular}
\end{center}
\end{table*}

 Finally, we give two examples to compare the
computation load of our new strategy and the original strategy, by
two concrete values $k=2$ and $k=3$. It can be seen our
new repair strategy needs much less computation.

\begin{Example}\label{Exm_computation_k_2}
When $k=2$, for the $(4,2)$ Hadamard MSR code determined by the
coding matrices given in Example \ref{Exm_1}, the computation load
is given in Table 3.
\begin{table*}
\begin{center}
\caption{computation load of $(4,2)$ Hadamard MSR code}
\begin{tabular}{|c|c|c|c|}
\hline
Node & Repair & & \\
to repair & strategy & \rb{ADD} & \rb{MUL}\\
\hline
Systematic & New & 28 & 17\\
\cline{2-4}
node 1 & Original & 132& 28 \\
\hline
Systematic & New & 28& 17 \\
\cline{2-4}
node 2 & Original & 132& 28 \\
\hline
Parity & New & 28& 15 \\
\cline{2-4}
node 1 & Original & 132 &24 \\
\hline
Parity & New  &  28 & 20\\
\cline{2-4}
node 2 & Original  & 152& 120 \\
\hline
\end{tabular}
\end{center}
\end{table*}
\end{Example}

\begin{Example}\label{Exm_computation_k_3}
When $k=3$, for the $(5,3)$ Hadamard MSR code determined by the
coding matrices given in Example \ref{Exm_2}, the computation load
is given in Table 4.
\begin{table*}
\begin{center}
\caption{computation load of $(5,3)$ Hadamard MSR code}
\begin{tabular}{|c|c|c|c|}
\hline
Node & Repair &  & \\
to repair & strategy & \rb{ADD} & \rb{MUL} \\
\hline
Systematic & New  & 80 & 42 \\
\cline{2-4}
node 1 & Original  & 528 & 128 \\
\hline
Systematic & New &  80 & 42\\
\cline{2-4}
node 2 & Original & 528 & 128 \\
\hline
Systematic & New  & 80 & 28 \\
\cline{2-4}
node 3 & Original & 528 & 256 \\
\hline
Parity & New & 80 & 44 \\
\cline{2-4}
node 1 & Original  & 528 & 272 \\
\hline
Parity & New & 80 & 66 \\
\cline{2-4}
node 2 & Original  & 736 & 576 \\
\hline
\end{tabular}
\end{center}
\end{table*}
\end{Example}


\section{Conclusion}\label{section_of_conclusion}

In this paper, a new repair strategy of Hadamard MSR code was presented, which can be regarded as a generalization of the original repair strategy. By choosing the standard basis, our strategy can dramatically decrease the computation load in contrast to the original one.

\end{document}